\documentclass[
   ,final
  ]
  {aipproc}

\layoutstyle{8x11double}

\SetInternalRegister\hbadness{8000} 

\newcommand\doingARLO[2][]{%
  \ifx\mmref\undefined #1\else #2\fi
}

\begin{document}

\title 
      [Infandum, regina, iubes renovare dolorem]
      {The Connection Between Pulsation, Mass Loss and Circumstellar Shells in Classical Cepheids}

\classification{97.10.Ex,97.10.Fy,97.10Me, 97.10.Sj, 97.30.Gj}
\keywords{Cepheids, Circumstellar Matter, Stars: Mass Loss}

\author{Hilding Neilson}{
  address={Department of Astronomy and Astrophysics, University of Toronto},
  email={neilson@astro.utoronto.ca}
}

\iftrue
\author{Chow-Choong Ngeow}{
  address={University of Illinois at Urbana-Champaign}  
}

\author{Shashi Kanbur}{
  address={State University of New York at Oswego}
}

\author{John B. Lester}{
  address={University of Toronto Mississauga}
}
\fi

\copyrightyear  {2009}

\begin{abstract}
Recent observations of Cepheids using infrared interferometry and Spitzer photometry have detected the presence of circumstellar envelopes (CSE) of dust and it has been hypothesized that the CSE's are due to dust forming in a Cepheid wind. Here we use a modified Castor, Abbott \& Klein formalism to produce a Cepheid wind, and this is used to estimate the contribution of mass loss to the Cepheid mass discrepancy Furthermore, we test the OGLE-III Classical Cepheids using the IR fluxes from the SAGE survey to determine if Large Magellanic Cloud Cepheids have CSE's.  It is found that IR excess is a common phenomenon for LMC Cepheids and that the resulting mass-loss rates can explain at least a fraction of the Cepheid mass discrepancy, depending on the assumed dust-to-gas ratio in the wind.
\end{abstract}

\date{\today}

\maketitle
\section{Introduction}
Cepheid variable stars have been studied for more than 200 years, since the discovery of variability in $\delta$ Cephei \citep{Goodricke1786}.  Even now, there are still many unanswered questions about the structure and evolution of these stars and new questions are being raised.  One such challenge is the discovery of circumstellar envelopes surrounding Galactic Cepheids using infrared interferometry \cite{Kervella2006, Merand2006, Merand2007}.  

These envelopes appear as an IR excess which has been observed before.  Galactic Cepheids have been observed using IRAS and IR excess was detected in some Cepheids \citep{McAlary1986,Deasy1988}.  More recently,  Spitzer observations have confirmed IR excess around Galactic Cepheids \citep{Marengo2009}.  There is evidence for IR excess about $l$ Car and RS Pup based on observations of the spectral energy distribution from V-band to $100~\mu m$ \citep[also see Alexandre Gallenne's poster]{Kervella2009}.  These observations provide compelling evidence for the existence of circumstellar envelopes surrounding Cepheids.  

While observational evidence is mounting for the existence of the circumstellar envelopes, there are a number of possible explanations for what they are.  One possibility is that the envelopes are not envelopes at all but are instead disks \citep{Feast2008}.   This model was suggested as an alternative model for computing the distance to RS Pup from light echoes instead of  the spherically symmetric model \citep{Kervella2008}.  The disk model was proposed as an alternative solution only and was not intended to necessarily be a physical explanation, however, it is important to consider the model.  One argument against the disk model is the fact that interferometry has detected shells about every Cepheid that has been observed.  If the IR excess were due to dusty disks then one would expect that the amount of excess would depend on the inclination of the disk which is random with respect to the line of sight of the observer.  Hence one would expect only a small fraction of Cepheids to exhibit an IR excess, contrary to what has been observed.

Another possibility is that the envelopes are relics from earlier stages of stellar evolution.  For instance, it is argued that the cold component of the nebula surrounding RS Pup is interstellar in nature \citep{Kervella2009}, and the argument is supported by observations of a circumstellar envelope surrounding the Herbig Be star HD 200775.  The Be star may be an analog to an earlier stage of evolution for RS Pup. The connection is strengthened by the observed elongation of the nebula, suggesting bipolar mass loss \citep{Bond2009}.  However, there is no evidence that the warm circumstellar envelopes are evolutionary relics, and interferometric observations of the non-pulsating yellow supergiant $\alpha$ Persei suggest that the CSE's are related to the evolution of Cepheids.

A simple explanation for the existence of CSE's is mass loss from the Cepheids themselves.  This is not a new idea,  Cepheid mass loss has been modeled and it was found that Cepheids could lose about $10^{-7}~M_\odot/yr$. This amount of mass loss would have a significant effect on the evolution of a Cepheid \citep{Brunish1987}.  The concept of Cepheid mass loss has been ``rediscovered'' with the observations of CSE's.  Furthermore, spectroscopic observations of a number Cepheids found asymmetry in the H$\alpha$ line consistent with an outflow \citep{Nardetto2008}.

If the CSE's are generated from Cepheid mass loss then this mass loss affects the IR Leavitt Law (Period-Luminosity relation), may add an additional uncertainty to the IR surface technique for determining the angular diameter of Cepheids, and play a role in the evolution of these stars.  It is important to characterize the nature of the CSE's and the mass-loss mechanism of Cepheids.   One approach to understanding CSE's is to model the IR excess of a large sample of Cepheids.  Fortunately, there exists IR observations of a large number of Cepheids thanks to the correlation of the OGLE-III Cepheids \citep{Soszynski2008} with the SAGE survey of the Large Magellanic Cloud \citep{Meixner2006} as well as the 2MASS survey \citep{Ngeow2009}.  This correlation yields VIJHK and IRAC magnitudes for about $1800$ Cepheids for each of the two epochs of SAGE observation plus the average flux from the two epochs.  Using this data we can test the existence of CSE's in LMC Cepheids and if the source of the CSE's is mass loss.

\section{Method}
In this work, we test the existence of CSE's surrounding LMC Cepheids by fitting the spectral energy distribution of each Cepheid in the sample.  The brightness as a function of wavelength is fit assuming a stellar luminosity plus a dust shell luminosity, where the dust forms at a larger distance from a Cepheid and is due to mass loss.  The stellar luminosity is 
\begin{equation}
L_\nu(Star) = 4\pi R_*^2 \pi B_\nu(T_{\rm{eff}}),
\end{equation}
and the shell luminosity is
\begin{eqnarray}
&& \nonumber L_\nu(Shell) = \frac{3}{4\pi}\frac{<a^2>}{<a^3>}\frac{\dot{M}_{dust}}{\bar{\rho} v_{dust}} \\ &&  \times Q_\nu^A \int_{R_c}^\infty B_\nu(T_{dust})[1-W(r)]dr.
\end{eqnarray}
The shell luminosity depends on the mean surface area and volume of the dust grains as well as the mean density of the dust grain.  The dust is assumed to be silicate dust with a mean density of $\bar{\rho} = 3.7~g/cm^3$, the dust ranges in size from $a = 0.005$ to $ 0.25~\mu m$ and we use the Mathis et al. dust size distribution \citep{Mathis1977} for computing $<a^2>$ and $<a^3>$.  The dust velocity is about $100~km/s$ and the absorption coefficient $Q_\nu^A \propto 1/\lambda$ for $\lambda < 10~\mu m $.  The term $W(r)$ is the dilution factor and is proportional to $r^{-1/2}$.  The inner boundary is the condensation radius, defined as the distance from the star where the gas reaches a temperature $T = 1200~K$.  The dust is assumed to be in equilibrium with the stellar radiation and hence the dust temperature $T_{dust} = T_{\rm{eff}}W(r)^{1/4}$ \citep{Ivezic1997}.
\begin{figure}[t]
\label{f2}
\caption{(Left) The values of $\chi^2$ for the fit of the radius and mass-loss rate for each Cepheid in the OGLE-III sample for each epoch of SAGE observation as a function of period.  The horizontal line represents the cut-off of $\chi^2 = 1.25$.  (Right) The best-fit radius of each Cepheid where the value of $\chi^2< 1.25$.}
\includegraphics[width=0.5\textwidth]{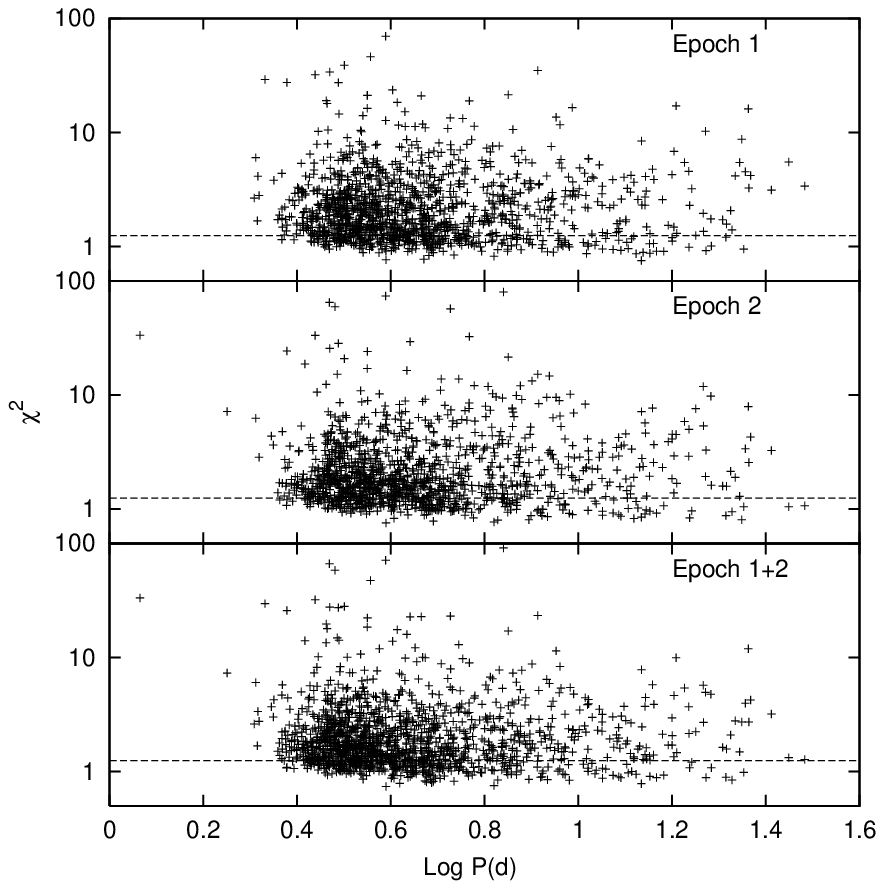}\includegraphics[width=.5\textwidth]{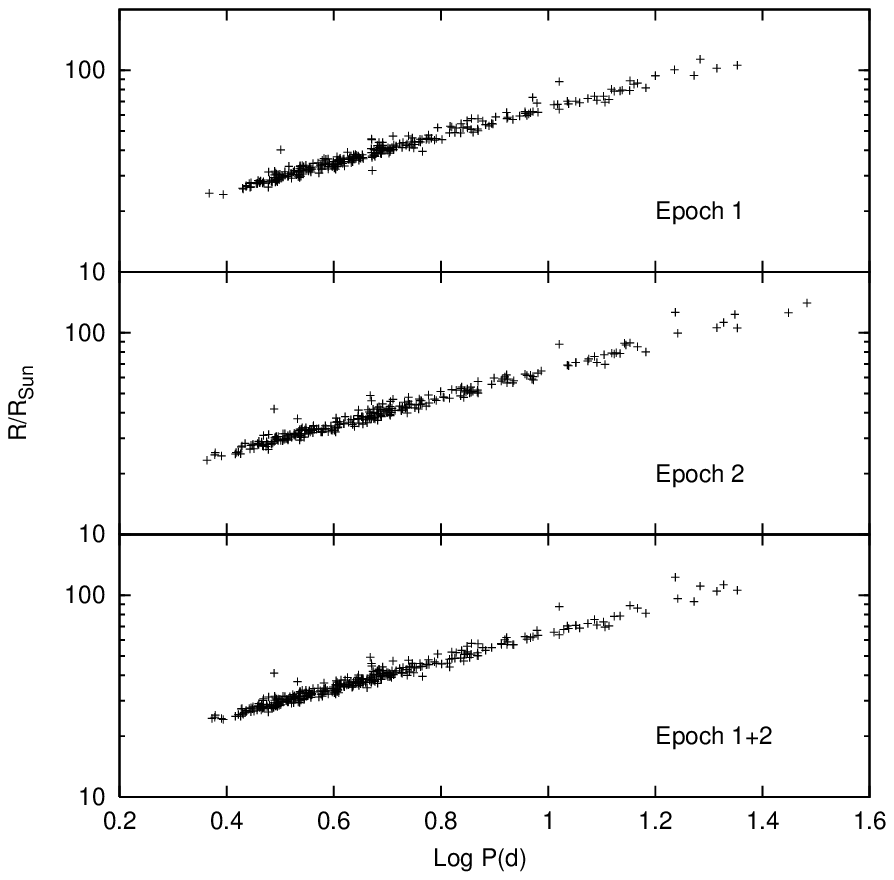}
\end{figure}

The stellar temperature is estimated using a temperature-color-period relation for LMC Cepheids \cite{Beaulieu2001}.  Thus if we can guess the radius and the mass-loss rate of a Cepheid then we can compute the total luminosity (sum of stellar plus shell) and assuming a distance modulus of $18.5$, we predict the apparent magnitudes of each Cepheid as a function of wavelength.  Therefore we $\chi^2$-fit these two parameters for each Cepheid in the sample and predict radii and mass-loss rates.

\section{Results}
\begin{figure}[t]
\label{f3}
\caption{The best-fit mass-loss rates for the LMC Cepheids from the OGLE-III survey with $\chi^2<1.25$. for each epoch of SAGE observation.}
\includegraphics[width=0.75\textwidth]{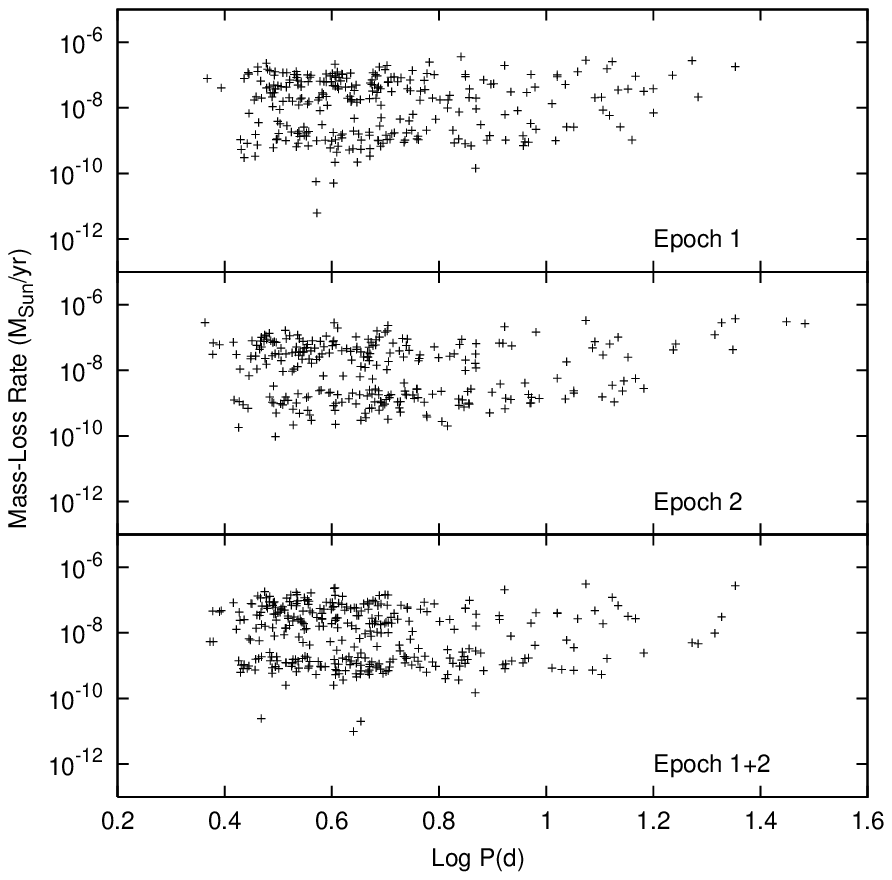}
\end{figure}

The values of the $\chi^2$ fit for each epoch of SAGE observation is shown in Figure \ref{f2}. Because of the large spread of values we apply a $\chi^2$-cut of $1.25$ to the data.  Also, one might argue that the because of the large spread of $\chi^2$ values that mass loss is not a reasonable model but we have computed $\chi^2$ fits of radius only and find a similar spread of fits.  The predicted radii and mass-loss rates of the Cepheids with $\chi^2 < 1.25$ are shown in Figures \ref{f2} and \ref{f3}.   It is worth noting that fitting a Period-Radius (P-R) relation to the predictions yields a slope of $0.68$ consistent with observed P-R relations.  The mass-loss rates range from about $10^{-11}$ to $10^{-7}~M_\odot/yr$ suggesting that mass loss is an important phenomena in Cepheids. 

\begin{figure}[t]
\label{f4}
\caption{(Left) The predicted radiative-driven mass-loss rates for Galactic Cepheids computed using the CAK method. (Right) The predicted pulsation/shock-driven mass-loss rates for the same sample of Cepheids.  The dotted line represents the relation between the radiative-driven mass-loss rates and the pulsation period.}
\includegraphics[width=.5\textwidth]{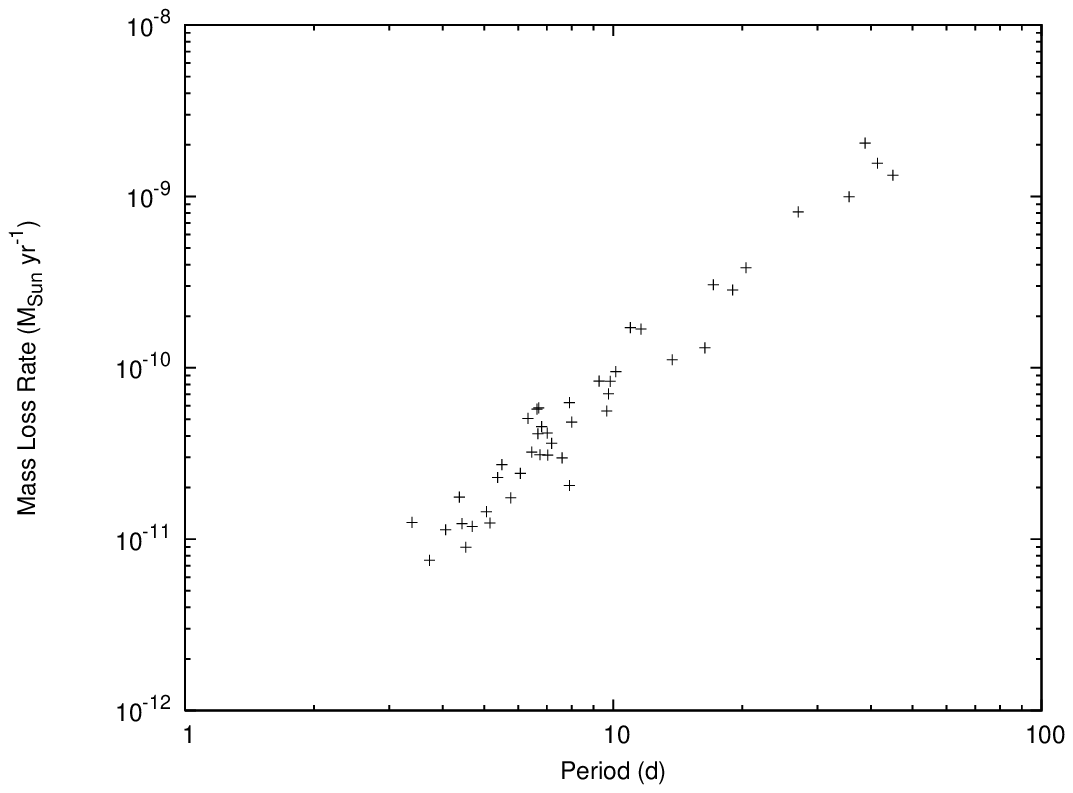}\includegraphics[width=.5\textwidth]{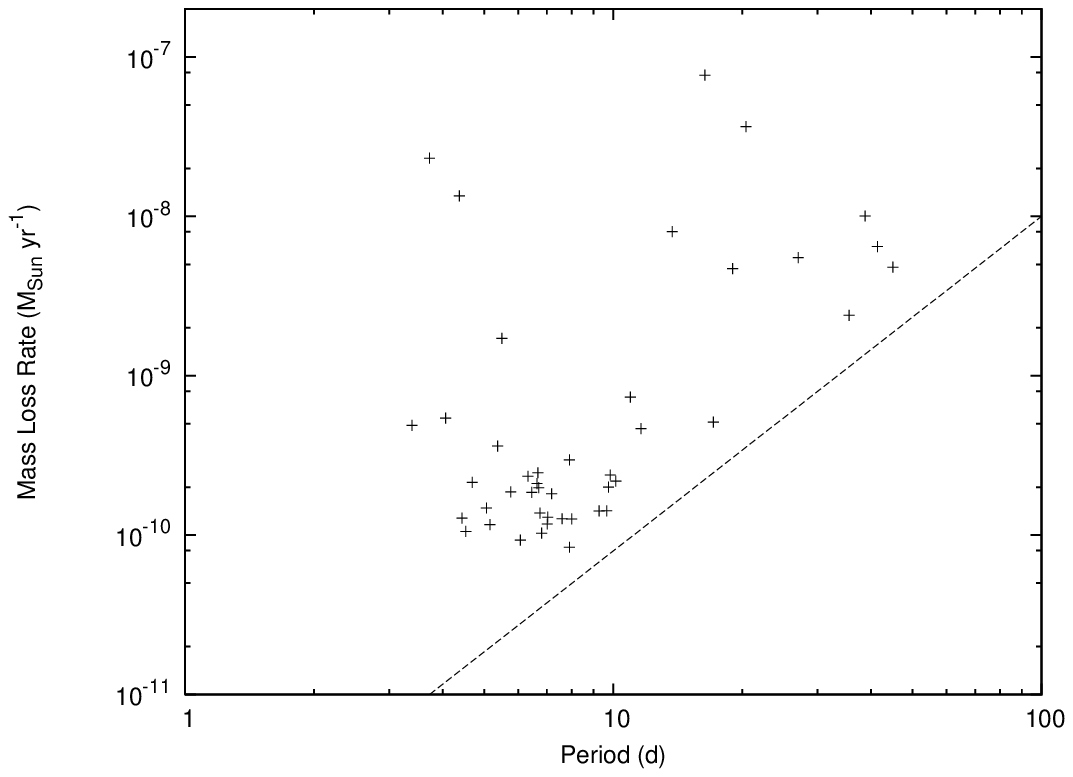}
\end{figure}

One question is what is the driving mechanism for mass loss in Cepheids.  In earlier works \citep{Neilson2008,Neilson2009}, the mechanism was explored using a sample of nearby Galactic Cepheids \citep{Moskalik2005}.  The Castor, Abbot and Klein (CAK) model for radiative-driven winds \citep{Castor1975} is applied to the sample and radiative-driven mass-loss rates are computed.  These rates are shown in Figure \ref{f4} and a visual comparison of the predicted radiative-driven mass-loss rates with the best-fit rates for the LMC Cepheids suggest that the mass loss cannot be due to radiative driving.  The mass-loss rates for short-period Cepheids are significantly smaller than the computed rates and there is an obvious correlation between the radiative-driven mass-loss rates and the pulsation period that is not seen in the LMC Cepheids.  

It can easily be concluded that radiative driving is an insufficient driving mechanism.  We have proposed an alternate theory that mass loss is driven by pulsation and shocks generated in the atmosphere due to pulsation similar to earlier works \citep{Brunish1987}.  The idea is tested using modified version of the CAK method to include the effect of pulsation and shocks.  The shock velocities are taken from the hydrodynamic model of $\delta$ Cephei \citep{Fokin1996} with a scaling relation derived to compute shock velocities for other Cepheids.  The pulsation/shock-driven mass-loss rates for the sample of Galactic Cepheids is shown in Figure \ref{f4} as a function of pulsation period.  It is found that the pulsation/shock-driven mass-loss rates do not have a period dependence.   The mass-loss rates also range from $10^{-10}$ to $10^{-7}~M_\odot/yr$, although only a few Cepheids have large mass-loss rates.

We also compare predicted mass-loss rates from for theoretical models \citep{Bono2000} of Galactic, LMC, and SMC Cepheids  \citep{Neilson2009} to the computed mass-loss rates of the LMC Cepheids. The pulsation/shock-driven mass-loss rates are determined to be up to $2\times 10^{-7}~M_\odot/yr$ which is comparable to the mass-loss rates determined here from observations.   This is further evidence that the mass-loss mechanism for Cepheids is due to pulsation and shocks.

The pulsation/shock theory for Cepheid mass loss roughly agrees with behavior of the best-fit mass-loss rates for the LMC Cepheids as a function of period, but seems to predict smaller mass-loss rates. One reason the mass-loss rates may be underestimated is that the modified-CAK method requires knowledge of the mass of a Cepheid and this was done assuming a Period-Mass-Radius relation derived from adiabatic pulsation models.  It is likely these masses are overestimated and if the masses are actually smaller then the mass-loss rates may increase significantly.  Another source of error is that the mass-loss model is a quasi-static model and ignores time-dependent dynamics in the atmosphere of a Cepheid.

\section{Conclusions}
In this presentation, we have shown computed mass-loss rates for LMC Cepheids from the OGLE-III survey that have been correlated with the 2MASS and SAGE surveys to determine IR fluxes.  The mass-loss rates of Cepheids are found to range from $10^{-10}$ to $10^{-7}~M_\odot/yr$.  These mass-loss rates contribute a shell flux that is about $10-30\%$ of the stellar flux at $8.0~\mu m$.

These rates are not strictly period-dependent as seen for prediction of radiative-driving.  There is no obvious reason to expect that Cepheid mass loss should be a function of pulsation period, especially since a period dependence is equivalent to a Reimer's relation for Cepheid mass loss.  A Reimer's relation suggests that the potential energy of the wind is proportional to the stellar luminosity, however, shocks in the atmosphere are not necessarily proportional to the luminosity alone.  The shocks are found to be non-linearly proportional to the luminosity, pulsation period and mass \citep{Neilson2008}, which leads to a significant deviation from a Reimer's relation.  Therefore it is reasonable for the mass-loss rates to not be a strict function of period.

Mass loss is a reasonable explanation for the existence of the CSE's surrounding Cepheids and hence suggests that there is an IR excess for the majority of Cepheids.  This IR excess affects the structure of the IR Leavitt Law \citep{Ngeow2009, Ngeow2008, Freedman2008, Madore2009} by making Cepheids appear brighter than they actually are.  This result is not significant if the amount of IR excess is similar for all Cepheids independent of metallicity but if the IR excess depends on metallicity then mass loss will cause an additional uncertainty to the IR Leavitt Law for computing distances.

If the mass loss affects the IR Leavitt Law, then it is also reasonable to expect that mass loss will contribute to the IR surface brightness technique and potentially cause Cepheid angular diameters to be overestimated. It is important to characterize the how much angular diameters are overestimated if one wishes to compute distances to Cepheids using the Baade-Wesselink method.

Cepheid mass loss may also be a solution to the Cepheid mass discrepancy \citep{Caputo2005, Keller2008}, where there are two arguments for the current value of the mass discrepancy.  Caputo et al. \citep{Caputo2005} argues the mass discrepancy is about $20\%$ for small mass Cepheids ($M\approx 4~M_\odot$) and decreases as a function of mass. On the other hand, Keller \citep{Keller2008} argues the mass discrepancy is about $17\%$ for all masses.  If we consider the analysis in this work, short-period Cepheids $(\log P<1)$, with an average mass-loss rate of $10^{-7}~M_\odot/yr$ will lose about $1~M_\odot$ over a crossing of the instability strip which is consistent with both arguments for the mass discrepancy.  However, for long-period Cepheids the total mass lost over a crossing of the instability strip will a smaller fraction of the stellar mass because the evolutionary timescale decreases with increasing mass.  That result is consistent with the arguments of Caputo et al. but not of Keller.  It is necessary to better understand both Cepheid mass loss and the mass discrepancy to determine if mass loss is truly the solution for the mass discrepancy.

\doingARLO[\bibliographystyle{aipproc}]
          {\ifthenelse{\equal{\AIPcitestyleselect}{num}}
             {\bibliographystyle{arlonum}}
             {\bibliographystyle{arlobib}}
          }
\bibliography{sp}

\end{document}